\documentstyle[12pt,epsfig]{article}
\begin{document}
\title{Diffusion and drift of cosmic rays in highly turbulent magnetic fields}
\author{Juli\'an Candia$^{a,b}$ and Esteban Roulet$^c$\\
$^a${\small\it IFLP, Departamento de F\'{\i}sica, Universidad Nacional de La Plata, 
C.C. 67,}\\{\small\it La Plata 1900, Argentina}\\
$^b${\small\it Fermi National Accelerator Laboratory, P.O. Box 500,}\\
{\small\it Batavia, IL 60510, USA}\\
$^c$ {\small\it CONICET, Centro At\'omico Bariloche, Av. Bustillo 9500,}\\
{\small\it Bariloche 8400, Argentina}}
\maketitle
\vspace*{-11.4cm}
\noindent \makebox[10cm][l]{\small \hspace*{-.2cm} } {\small
FERMILAB-Pub-04/141-T} \\
\makebox[10cm][l]{\small \hspace*{-.2cm} } {\small astro-ph/0408054 } {} \\
{\small } \\
\vspace*{9.cm}
\begin{abstract}
We determine numerically the parallel, perpendicular, and antisymmetric diffusion coefficients
for charged particles propagating in highly turbulent magnetic fields, by means of extensive Monte
Carlo simulations. We propose simple expressions, given in terms of a small set of fitting parameters,
to account for the diffusion coefficients as functions of magnetic rigidity and turbulence level, and corresponding to 
different kinds of turbulence spectra. The results obtained satisfy scaling relations, which make them useful for describing 
the cosmic ray origin and transport in a variety of different astrophysical environments. 
\end{abstract}
The diffusion and drift of charged particles across highly turbulent magnetic fields are
key issues in describing the transport of cosmic rays in different astrophysical environments, e.g. 
the interplanetary, interstellar and intergalactic media, as well as the efficiency of Fermi 
acceleration processes at cosmic ray sources. In particular, it has been shown that the inclusion
of drift effects in the transport equation leads naturally to an explanation for the knee, for the
second knee and for the observed behavior of the composition and anisotropies between the knee and
the ankle \cite{pt93,ca02a,ca02b,ca03}. However, the accuracy of the investigations performed so far are limited
by the lack of conclusive results concerning the behavior of the diffusion tensor under highly
turbulent conditions as a function of the particle energy and the relevant magnetic field parameters.   
In particular, the magnetic fields in the Galaxy are highly turbulent because 
 the mean random field 
is of the order of the mean regular field. Moreover, since 
there are reversals in the orientation
of the regular field, this implies the existence of regions with negligible
 regular fields in which the turbulence prevails.  

Perturbative studies for low turbulence have been developed since long ago \cite{pa65,jo66,ch70}, but these analytic methods 
cease to be applicable for high turbulence levels, and only recently 
the parallel and transverse diffusion coefficients were calculated numerically for regimes with
high turbulence \cite{gi99,ca02}. The aim of this work is to provide a thorough and more systematic calculation
of these coefficients, and to parametrize the results in order to make them useful in a variety of different
kinds of applications. Moreover, we present here also a numerical evaluation of the Hall diffusion coefficient
that is responsible for the drift effects, which so far has never been evaluated quantitatively under highly
turbulent conditions. It should also be remarked that, while in \cite{gi99,ca02} only the Kolmogorov spectrum of 
fluctuations in the random magnetic field was considered, 
in this work other types of turbulence  are studied as well
(namely, the Kraichnan and Bykov-Toptygin turbulence spectra, which bracket a wide range 
of possible turbulence spectra).    

A relativistic particle of charge $Ze$ propagating in an uniform regular magnetic field ${\bf B_0}$ describes a 
helical path characterized by a pitch angle $\theta$ and a Larmor radius given by 
\begin{equation}
r_L\equiv {pc\over ZeB_0}\simeq{{E/Z}\over{10^{15}~{\rm eV}}}\left({{B_0}
\over{\mu{\rm G}}}\right)^{-1} {\rm pc}\ .
\label{rl} 
\end{equation}
The component of the velocity parallel to ${\bf B_0}$ is $v_\parallel =c\cos\theta$, while the radius of the
helical trajectory is $r_L\sin\theta$.  In the presence of a random magnetic field ${\bf B_r}$ with a maximum
scale of turbulence $L_{max}$, 
the particles scatter off the magnetic irregularities and change their pitch angle, 
but not their velocity. 
The pitch angle scattering proceeds mainly in resonance 
(i.e.,  the scattering is dominated by 
the inhomogeneities with scales  of the order of $r_L$), and hence it 
is an effective mechanism of isotropization as long as 
$r_L< L_{max}$. For instance, for the galactic magnetic field, with strength $B_0\simeq$~few~$\mu$G and 
maximum scale of turbulence $L_{max}\simeq 100$~pc, the pitch angle scattering leads to a 
diffusive regime for protons with energies up to few~$10^{17}$~eV. 

In general, the diffusion tensor $D_{ij}$ can be written as
\begin{equation}
D_{ij}=\left(D_{\parallel}-D_{\perp}\right)b_ib_j+D_{\perp}\delta_{ij}+D_A\epsilon_{ijk}b_k
\end{equation}
where ${\bf{b}}={\bf{B_0}}/B_0$ is a unit vector along the regular magnetic field,
$\delta_{ij}$ is the Kronecker delta symbol, and $\epsilon_{ijk}$ 
is the Levi-Civita fully antisymmetric tensor.   
The symmetric terms contain the diffusion coefficients parallel and perpendicular to the
mean field, $D_\parallel$ and $D_\perp$, which describe diffusion due to small-scale turbulence, 
while the antisymmetric term contains the Hall diffusion coefficient $D_A$.

The diffusion along the magnetic field direction is due to the pitch 
angle scattering and leads to a diffusion 
coefficient given by 
\begin{equation}
D_\parallel={{c}\over{3}}\lambda_\parallel\ ,
\label{dpar1}
\end{equation}
where $\lambda_\parallel$ is the mean free path in the parallel direction \cite{pt93}. In this expression, 
$\lambda_\parallel$ depends on the power of the random magnetic field modes at scales of order of $r_L$, i.e.
\begin{equation} 
\lambda_\parallel\propto\left.{{r_L}\over{{\rm d}E_r/{\rm d\ ln}k}}\right|_{k=2\pi/r_L}\ ,
\label{dpar2} 
\end{equation} 
where ${\rm d}E_r/{\rm d}k$ is the power spectrum of the random magnetic field
energy density. 

In general, a component of the diffusion tensor is associated to some particular physical processes which may
not contribute to the other components. The diffusion transverse to the regular magnetic field is due to both 
pitch angle scattering, which is the mechanism that prevails in the parallel diffusion, and to the wandering of 
the magnetic field lines themselves, which drag with them the diffusing particles in the direction perpendicular to
${\bf B_0}$ \cite{jo66,jo69,fo74}. 
When the turbulence level is small, the diffusion in the orthogonal direction 
is much slower than in the 
parallel one and is strongly affected by the field line random walk, 
but in the limit of very high turbulence, the parallel and perpendicular motions become similar.  

The Hall coefficient $D_A$ describes in turn the macroscopic drift associated to the gradient of the CR
density, leading to a current 
\begin{equation}  
{\bf J_A}=D_A{\bf b}\times\nabla N\ .
\label{jA}
\end{equation}  
This macroscopic drift is orthogonal to both ${\bf B_0}$ and $\nabla N$, and
in particular is also present for a constant regular field.
The relation between this macroscopic current and the microscopic drift associated to gradients and curvature in ${\bf B_0}$
has led to some confusion and controversy in the literature \cite{sp62,no63,ro70,bu85}. 
A common approach is to identify a guiding center, associated to the instantaneous radius of curvature of the particle's 
trajectory, and to define a guiding center velocity ${\bf v_g}$. In the so-called first order orbit theory, an average 
over a particle's gyroperiod is performed assuming that the scale of variations in the field is much smaller than the 
Larmor radius, and then an ensemble average over a given distribution of particles is carried out. Alternatively, 
$\langle{\bf v_g}\rangle$ can be calculated by averaging locally over a distribution of particles in phase space \cite{bu85}.
In the latter case, the results are given in terms of an expansion of the anisotropy of the momenta 
in spherical harmonics, but in principle no assumptions concerning the scale of variations of the field are required. 
On the other hand, the average 
particle velocity $\langle{\bf v}\rangle$ can be calculated from the Vlasov (or collisionless Boltzmann) equation, 
averaging again over a particle distribution in phase space. As shown in \cite{bu85}, the mean particle and guiding 
center velocities are related through their definition, and they differ in a term which is sometimes called diamagnetic drift.
A qualitative difference that can be pointed out is that $\langle{\bf v}\rangle$ has a contribution arising from density 
gradients and can be nonvanishing even in the case of an uniform regular magnetic field, while $\langle{\bf v_g}\rangle$
depends only on the field gradients. 

Notice that, when making an average of the particle velocities inside a given volume element, one is including 
particles whose center of gyration are outside the volume considered, and, inversely, considering the guiding
centers inside a given volume may correspond to taking into account particles which are actually outside that
volume element. Since the observable quantity is the average particle velocity $\langle{\bf v}\rangle$ and not
the abstract concept of the guiding center motion $\langle{\bf v_g}\rangle$, it is clear that it is the former
one that will enter into the drift motions.  
  
As commented above, the diffusion coefficients were calculated analytically only 
in the case of small turbulence levels,      
while results valid under highly turbulent conditions require instead a numerical approach \cite{gi99,ca02}. 
Following \cite{gi99}, we will consider charged particles propagating in a magnetic field of the form 
${\bf B}({\bf r})=B_0{\bf\hat{z}}+{\bf B_r}({\bf r})$, where the first term represents an uniform regular field directed 
along the $z-$direction, while the second corresponds to the random component. 
In order to approximate numerically the isotropic and spatially homogeneous turbulent field, one can sum over a large 
number ($N_m$) of plane waves with wave vector direction, polarization and phase chosen randomly \cite{ba60,gi99}, i.e.
\begin{equation}
{\bf B_r}({\bf r})=\sum_{n=1}^{N_m}\sum_{\alpha=1}^2A(k_n){\bf\hat{\xi}}_n^\alpha
\cos({\bf k}_n\cdot{\bf r}+\phi_n^\alpha)\ ,
\label{bfourier}
\end{equation}    
where the two orthogonal polarizations ${\bf\hat{\xi}}_n^\alpha$ ($\alpha=1,2$) are in the plane perpendicular to the 
wave vector direction (i.e., ${\bf\hat{\xi}}_n^\alpha\perp{\bf k}_n$, so as to ensure that $\nabla\cdot{\bf B}=0$). 
The wavenumber distribution is taken according to a constant logarithmic spacing
between $k_{min}=2\pi/L_{max}$ and $k_{max}=2\pi/L_{min}$, 
where $L_{min}$ and $L_{max}$ are the minimum and maximum scales of turbulence, 
respectively. The energy density of the 
turbulent component is taken as ${\rm d}E_r/{\rm d}k\propto k^{-\gamma}$, 
where the spectral index $\gamma$ is given by the kind of mechanism that builds up the turbulence. 
Hence, the plane wave amplitudes satisfy 
$A^2(k_n)={\mathcal{N}}\langle B_r^2\rangle k_n^{-\gamma}(k_n-k_{n-1})$, with ${\mathcal{N}}$ a 
normalization constant that assures that $\sum_nA^2(k_n)=\langle B_r^2\rangle$.
In this work we will consider in particular three spectra of interest for astrophysical applications, 
namely a Kolmogorov spectrum with $\gamma=5/3$ (which is the only case studied numerically in the past, 
and is particularly attractive since, 
according to observations \cite{ar81,ru88}, the density fluctuations in the interstellar medium follow
this turbulence spectrum), a Kraichnan hydromagnetic spectrum with $\gamma=3/2$ \cite{kr65}, 
and the Bykov-Toptygin spectrum with $\gamma=2$ \cite{by87}.  
  
By means of the Kubo formalism \cite{ku57,fo77,bi97}, the diffusion coefficients $D_{ij}$ 
can be computed directly by  
taking ensemble averages of the decorrelation between different components of the single particle velocities, i.e.
\begin{equation}
D_{ij}=\int_0^\infty{\rm{d}}tR_{ij}(t) ,
\label{kubo}
\end{equation}
with $R_{ij}(t)=\langle v_{0i}v_j(t)\rangle$ (with $v_{0i}\equiv v_i(t=0)$ and 
where $\langle ...\rangle$ denotes the ensemble average taken over an 
isotropic distribution of many particles). 
In \cite{bi97}, it was assumed that the velocity decorrelations were 
modulated by exponential factors, adopting the following ans\"atze 
\begin{equation}
R_{xx}(t)=R_{yy}(t)={{c^2}\over{3}}\cos\omega t\ e^{-t/\tau_\perp}\ ,
\label{rxx}
\end{equation}
\begin{equation}
R_{yx}(t)=-R_{xy}(t)={{c^2}\over{3}}\sin\omega t\ e^{-t/\tau_A}\ 
\label{rxy}
\end{equation}
and
\begin{equation}
R_{zz}(t)={{c^2}\over{3}}\ e^{-t/\tau_\parallel}\ ,
\label{rzz}
\end{equation}
where $\omega=c/r_L$ is the Larmor angular gyrofrequency, 
and where $\tau_\parallel, \tau_\perp$ and $\tau_A$ are the decorrelation timescales associated to the different 
diffusion components. Then, the diffusion coefficients were obtained by integrating the proposed expressions 
for $R_{ij}(t)$ using Eq.(\ref{kubo}). 
The diffusion coefficients were found to be
\begin{equation}
D_\parallel={{cr_L}\over{3}}{\omega\tau_\parallel}\ ,
\label{dpar}
\end{equation}
\begin{equation}
D_\perp={{cr_L}\over{3}}{{\omega\tau_\perp}\over{1+\left(\omega\tau_\perp\right)^2}}
\label{dperp}
\end{equation}
and
\begin{equation}
D_A={{cr_L}\over{3}}{{\left(\omega\tau_A\right)^2}
\over{1+\left(\omega\tau_A\right)^2}}\ .
\label{da}
\end{equation}
It is interesting to note that expressions
of this form, but in which only a single timescale $\tau$ appears for the three diffusion coefficients, 
were obtained also in other analytic approaches that assumed a single scattering process to be responsible for
all the decorrelations \cite{gl69,is79,ba94}. 
However, it should be pointed out that the expressions proposed in \cite{bi97} for $R_{ij}(t)$ assume implicitly
a small departure from the helical trajectories, and they are no longer adequate for high turbulence levels 
(see below for further discussions). Moreover,
there is no general theory providing the decorrelation timescales, and this requires then further assumptions in 
order to make Eqs.(\ref{dpar})--(\ref{da}) useful. These additional assumptions are sometimes obscure and lead 
to results often at odds with the outcome of numerical simulations, as pointed out in \cite{ca02a}.

Alternatively, the parallel and perpendicular diffusion coefficients can
be calculated from the asymptotic rate of increase of the mean squared displacements in each direction,
namely    
\begin{equation}
D_\parallel=\lim_{\Delta t\to\infty}{{\langle(\Delta z)^2\rangle}\over{2\Delta t}}
\label{parplat}
\end{equation}
and  
\begin{equation}
D_\perp=\lim_{\Delta t\to\infty}{{\langle(\Delta x)^2\rangle}\over{2\Delta t}}=
\lim_{\Delta t\to\infty}{{\langle(\Delta y)^2\rangle}\over{2\Delta t}}\ .
\label{perpplat}
\end{equation}

In any case, the straightforward method consists basically in generating a sample configuration for the random magnetic
field (by choosing randomly the propagation direction, polarization and phase of the $N_m$ plane waves 
in Eq.(\ref{bfourier})),
and then following the trajectory of a particle that propagates in the total (regular plus random) field from the origin 
with a random initial direction. The results should then be averaged over a large number of 
different field configurations in
order to calculate the corresponding diffusion coefficients. In \cite{gi99}, for instance, the results were 
obtained by integrating firstly the trajectories of 2500 particles for a given field configuration, 
and then by averaging   
over 50 different field realizations. We found, however, that the results
are more biased by the particular 
choice of field configuration than by the choice of the initial 
velocity of the particle. Hence, it turns out to be
more convenient to average directly over a large number of field realizations. Indeed, we typically generated $\sim 10^5$ field 
configurations, following the trajectory of a single particle in each of them, 
and used $N_m=100$ modes in all simulations. 

For the numerical integration of the particle trajectories under the influence of the Lorentz force, 
we adopted a time step $\Delta t=0.1\ r_L/c$ in the Runge-Kutta routine. 
In \cite{ca02}, the parallel and perpendicular diffusion coefficients were computed from 
Eqs.(\ref{parplat})--(\ref{perpplat}), and this method requires to follow the particle trajectories for quite long 
times in order to reach the asymptotic region, typically longer than $t=1000\ r_L/c$. 
We found more convenient instead
to compute directly the particle decorrelation functions $R_{ij}(t)$ and integrate them through Eq.(\ref{kubo}), 
since this procedure requires to follow the particle trajectories for times typically not larger than $t=100\ r_L/c$. The underlying reason for this seems to be that the displacements keep more memory of the initial velocity adopted than the velocities themselves.  
In addition, this method allows to compute the antisymmetric diffusion coefficient by means of $R_{yx}$, while 
averages such as $\langle\Delta x\Delta y\rangle$ vanish for large times, and hence they are not useful for
computing $D_A$. 

Figure 1 shows the time dependence of the decorrelation functions associated to different velocity 
components, for the Kolmogorov case and different turbulence levels, with 
the turbulence level defined here\footnote{Alternatively, one can define the turbulence
level as $\eta\equiv\langle B_r^2\rangle/(\langle B_r^2\rangle+B_0^2)=\sigma^2/(1+\sigma^2)$ \cite{ca02}. The definition 
adopted here follows that in \cite{gi99}.}
 by $\sigma^2\equiv\langle B_r^2\rangle/B_0^2$. 
\begin{figure}[th!]
\centerline{{\epsfxsize=5.truein\epsfysize=3.5truein\epsffile{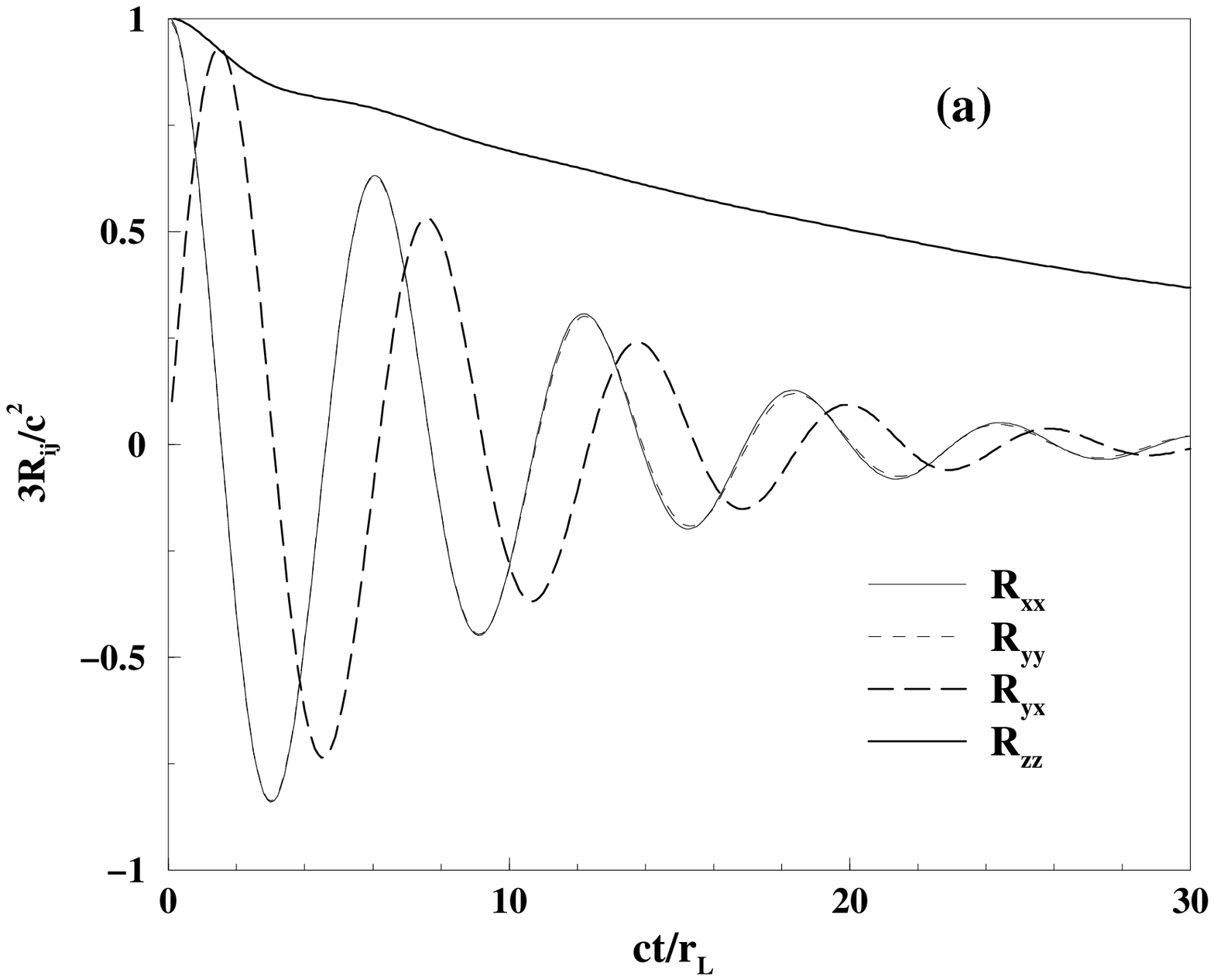}}}
\centerline{{\epsfxsize=5.truein\epsfysize=3.5truein\epsffile{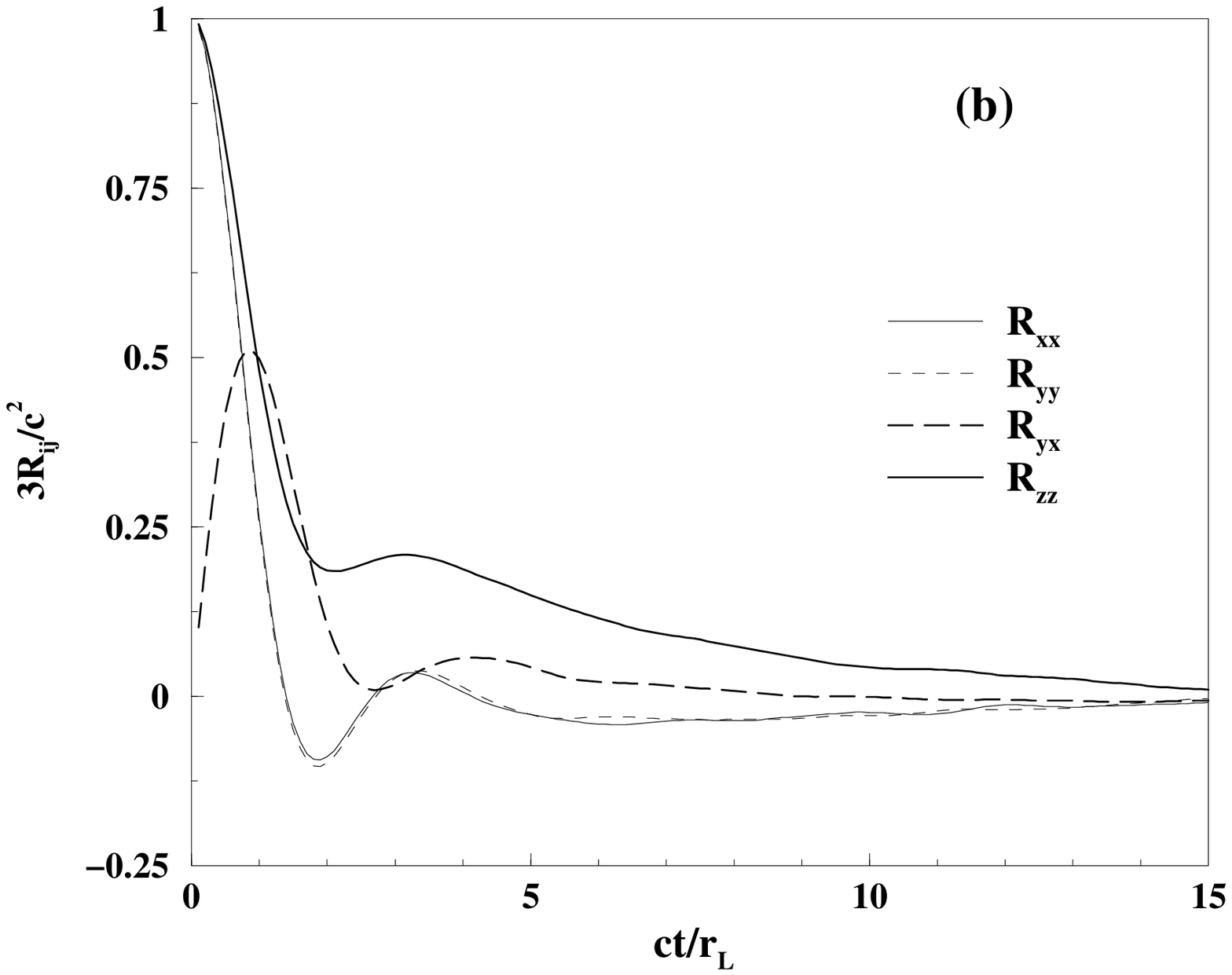}}}
\caption{Time dependence of the particle decorrelation functions associated to different velocity components, 
for the Kolmogorov turbulence spectrum and two different turbulence levels: (a) $\sigma^2=0.3$ and 
(b) $\sigma^2=5$, both corresponding to $r_L=0.1\ L_{max}$.}
\label{fig1}
\end{figure}
For low turbulence, it is observed in Figure 1(a) that, after a given initial transient period, 
the correlation functions behave as Eqs.(\ref{rxx})--(\ref{rzz}), 
and hence one can calculate the decorrelation timescales directly. For instance, 
following the local maxima of the sinusoidal functions one finds $\tau_\perp=\tau_A$, as it is apparent
from Figure 2. In \cite{bi97}, these decorrelation timescales were considered to be equal as a
simplifying assumption, but it was suggested that new effects could arise in a general case with $\tau_\perp\neq\tau_A$.
Here we find that these decorrelation timescales are indeed equal, 
and actually this could be expected from the fact 
that the physical origin for the decorrelations is the same in both cases.  
For high turbulence levels, however, the amplitude decrease is much more abrupt, as it is shown in Figure 1(b).  
The decorrelation functions become vanishingly small already in the transient period,
and Eqs.(\ref{rxx})--(\ref{rzz}) are hence not valid anymore.  

\begin{figure}[t]
\centerline{{\epsfxsize=5.truein\epsfysize=3.5truein\epsffile{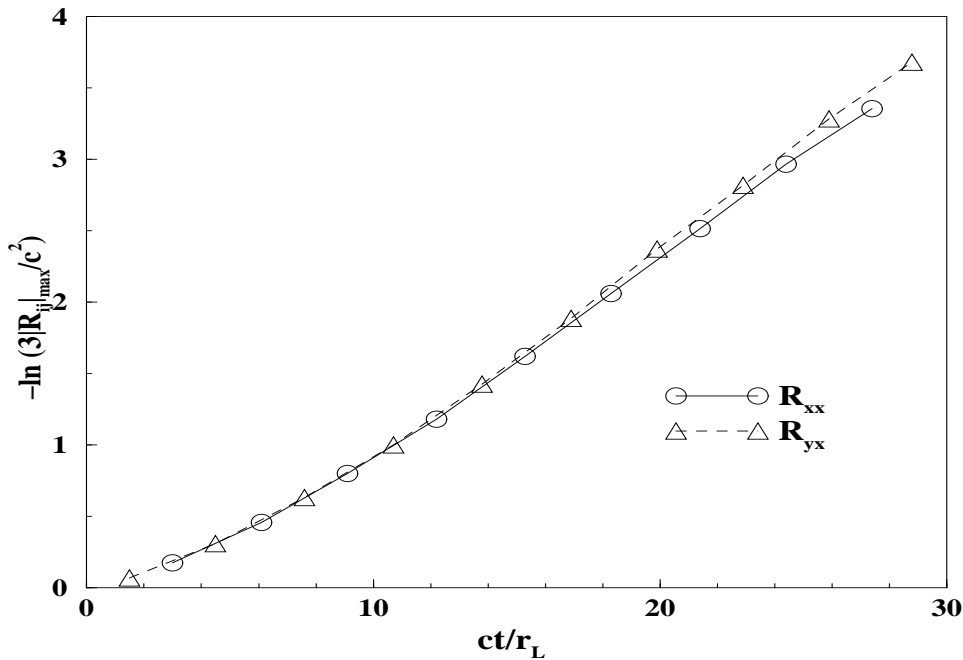}}}
\caption{Linear-log plots of the local maxima of $|R_{xx}|$ and $|R_{yx}|$ versus $t$
corresponding to the case of Fig.1(a), 
showing that $\tau_\perp=\tau_A$ (see Eqs.(\ref{rxx})--(\ref{rxy})).
The lines are guides to the eye.}
\label{fig2}
\end{figure} 

Defining the dimensionless parameter $\rho\equiv r_L/L_{max}$, 
the results given in terms of $D/(cL_{max}\rho)$ vs. $\rho$  
are universal and can be scaled in order to calculate the diffusion tensor for different sets of values 
for the regular field amplitude, random field length scale and particle energies. 
For instance, the range $0.01\leq\rho\leq 1$ investigated in this work can be regarded as corresponding to 
$10^{15}\leq E/\rm{eV}\leq 10^{17}$ for protons propagating in the Galaxy (for $B_0=1~\mu$G and $L_{max}=100$~pc).
Alternatively, considering protons propagating in the interplanetary field, with strength $B_0=50~\mu$G and 
$L_{max}=0.01$~AU, the limit to the diffusion regime at $\rho=1$ corresponds to the kinetic energy 
$E_k=1.8$~GeV.\footnote{In \cite{gi99}, however, the power spectrum of the random interplanetary field is 
considered to continue with constant amplitude from 0.01~AU up to a 
maximum scale of 1 AU, hence extending considerably the
diffusive regime to higher energies.} Since pitch angle scattering proceeds mainly in resonance, the particle propagation is essentially 
independent of the minimum scale length of turbulence $L_{min}$, as long as $L_{min}\ll r_L$. For definiteness, 
we then adopt $L_{min}=0.1\ r_L$.

\begin{figure}[t]
\centerline{{\epsfxsize=5.truein\epsfysize=3.5truein\epsffile{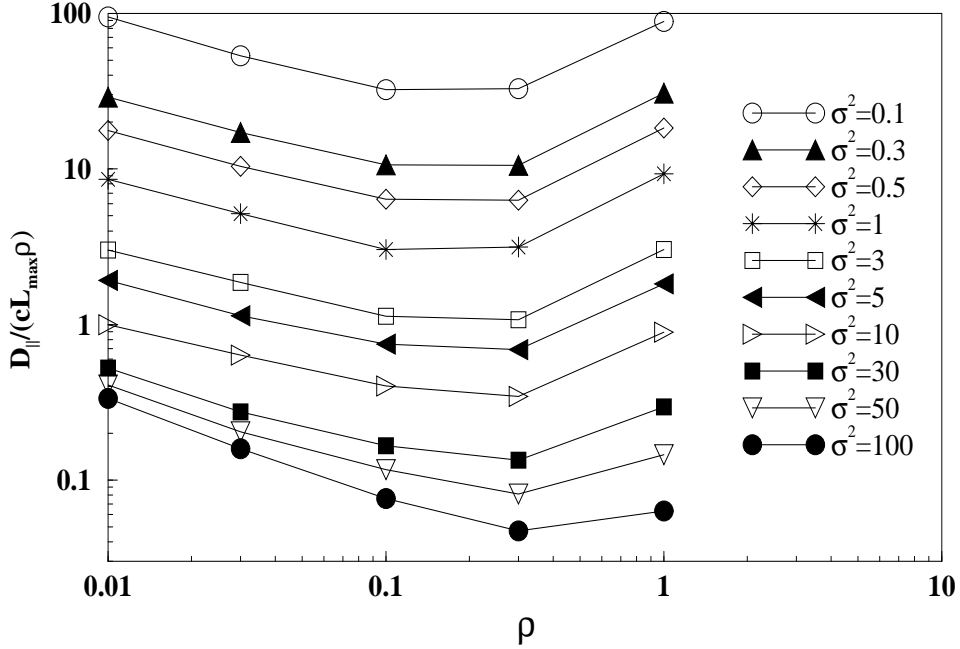}}}
\caption{Results obtained for the parallel diffusion and corresponding to the Kraichnan spectrum 
for different turbulence levels in the range $0.1\leq \sigma^2\leq 100$, as indicated. The lines are guides to the eye.}
\label{fig3}
\end{figure} 
Figure 3 shows the numerical results obtained for the parallel diffusion 
coefficient corresponding to the Kraichnan spectrum and 
for different turbulence levels in the range $0.1\leq \sigma^2\leq 100$, as indicated.  
As follows from Eqs.(\ref{dpar1})--(\ref{dpar2}), in the resonant scattering regime one expects that 
$D_\parallel\propto\rho^{2-\gamma}$, while at high rigidities (outside the resonant scattering regime) the 
scattering effectiveness is expected to decrease as $E^{-2}$ \cite{pt93,be90}, 
hence leading to $D_\parallel\propto\rho^2$, 
irrespective of the turbulence spectrum considered. The behavior of the results displayed in Figure 3 agrees 
very well with these expectations, showing a crossover between the resonant and non-resonant scattering regimes 
that takes place around $\rho\simeq 0.2$. Indeed, the actual scale length that separates both regimes is determined 
by the correlation length of the turbulence spectrum $L_c$, defined as 
\begin{equation}
\int_{-\infty}^\infty{\rm d}L\langle{\bf B_r}(0)\cdot{\bf B_r}({\bf r}(L))\rangle\equiv L_c\langle B_r^2\rangle\ ,
\end{equation}
where the point ${\bf r}(L)$ is displaced with respect to the origin by a distance $L$ along a fixed direction. 
Then, considering a spectrum of fluctuations of index $\gamma$ extending between the scale lengths $L_{min}$ and 
$L_{max}$, the correlation length is given by \cite{ha02} 
\begin{equation}
L_c={{1}\over{2}}L_{max}\ {{\gamma-1}\over{\gamma}}\ {{1-\left(L_{min}/L_{max}\right)^\gamma}\over
{1-\left(L_{min}/L_{max}\right)^{\gamma-1}}}\ .
\end{equation}
Hence, $L_c/L_{max}\simeq 0.2$ is a quite representative value for the random field power
spectra considered in this work, and this explains the change of regime observed at $\rho\simeq 0.2$. 

According to these considerations, an appropriate way of fitting the results is to interpolate $D_\parallel$ between the 
power laws that characterize the low- and high-rigidity regimes. 
A convenient way to achieve this is by means of the expression
\begin{equation}
{{D_\parallel}\over{cL_{max}\rho}}={{N_\parallel}\over{\sigma^2}}
\sqrt{\left({{\rho}\over{\rho_\parallel}}\right)^{2(1-\gamma)}+\left({{\rho}\over{\rho_\parallel}}\right)^2}\ ,
\end{equation}
where the parameters $N_\parallel$ and $\rho_\parallel$ are given in Table 1. 
This expression appears to fit well our numerical
results up to $\sigma^2\simeq 10$, which is already a very high turbulence level in most astrophysical 
applications of interest. For even higher turbulence levels, the results are best described asymptotically 
by the parameters $N_\parallel^{asint}$ and $\rho_\parallel^{asint}$, also given in Table 1. 

\begin{figure}[th!]
\centerline{{\epsfxsize=5.truein\epsfysize=3.5truein\epsffile{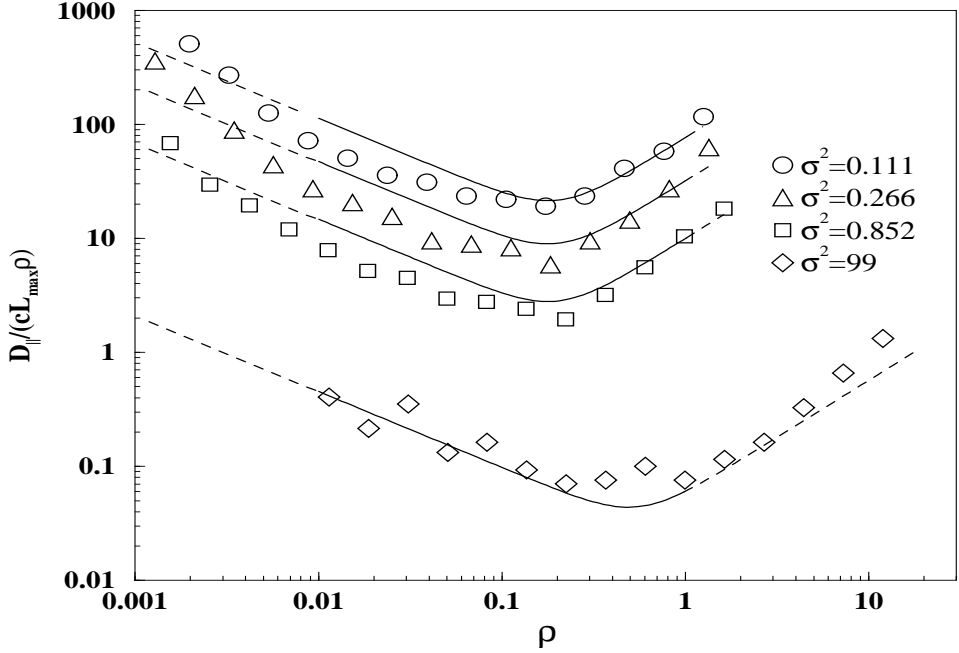}}}
\caption{Comparison between the fit to $D_\parallel$ given in this work 
(formally only valid in the range $0.01\leq\rho\leq 1$) and the results
of \cite{ca02}, both corresponding to a Kolmogorov spectrum
and for the same turbulence levels.} 
\label{fig4}
\end{figure} 

\begin{table}
\begin{tabular}{|r|r|r|r|r|r|r|r|r|} \hline
\multicolumn{1}{|c|}{Spectrum}&\multicolumn{1}{|c|}{$\gamma$}&\multicolumn{1}{|c|}{$N_\parallel$}&
\multicolumn{1}{|c|}{$\rho_\parallel$}&\multicolumn{1}{|c|}{$N_\parallel^{asint}$}&
\multicolumn{1}{|c|}{$\rho_\parallel^{asint}$}&\multicolumn{1}{|c|}{$N_\perp$}&\multicolumn{1}{|c|}{$a_\perp$}&
\multicolumn{1}{|c|}{$N_A$}\\ \hline
 Kraichnan & 3/2 & 2.0 & 0.22 & 3.5 & 0.65 & 0.019 & 1.37 & 17.6 \\
 Kolmogorov & 5/3 & 1.7 & 0.20 & 3.1 & 0.55 & 0.025 & 1.36 & 14.9 \\
 Bykov-Toptygin & 2 & 1.4 & 0.16 & 2.6 & 0.45 & 0.020 & 1.38 & 14.2 \\ \hline
\end{tabular}
\caption{Parameters of the fitting formulae corresponding to the parallel, transverse and antisymmetric diffusion
coefficients, and for different kinds of turbulence spectra. See more details in the text.}
\label{Table 1}
\end{table}

In order to check the consistency with the previous numerical results for highly turbulent parallel 
diffusion (which was studied in \cite{ca02} assuming a Kolmogorov spectrum of fluctuations), Figure 4 
shows a comparison between our fit and the results of \cite{ca02}, both corresponding to a Kolmogorov spectrum
and for the same turbulence levels. Notice that in \cite{ca02} the Larmor 
radius\footnote{To avoid any confusion, we adopt 
an asterisk to refer to the quantities defined in \cite{ca02}.}
$r_L^*$ is defined by replacing 
$B_0\to\sqrt{B_0^2+B_r^2}$ in Eq.(\ref{rl}), hence coupling in $r_L^*$ the dependence on rigidity and on the
turbulence level, while, on the other hand, the dimensionless parameter $\rho^*$ is defined as 
$\rho^*=2\pi r_L^*/L_{max}$. Hence, the results have to be rescaled according to the relations 
$D/(cL_{max}\rho)=D/(r_L^*c)/\sqrt{1+\sigma^2}$ and $\rho=\rho^*\sqrt{1+\sigma^2}/2\pi$. As can be seen
in Figure 4, both data sets agree reasonably well. 

\begin{figure}[t!]
\centerline{{\epsfxsize=5.truein\epsfysize=3.5truein\epsffile{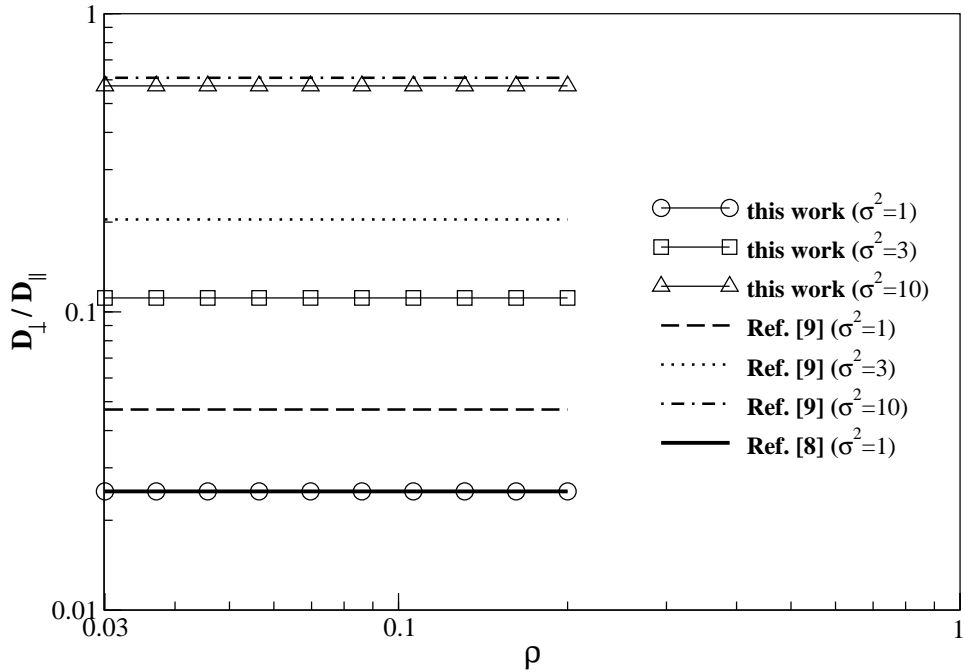}}}
\caption{Fit to $D_\perp/D_\parallel$ as a function of rigidity for the Kolmogorov spectrum and
corresponding to the low-rigidity regime (i.e. $\rho\leq 0.2$), 
as obtained for different turbulence levels. For larger rigidities, $D_\perp/D_\parallel\propto\rho^{-2}$ 
(see Eq.(\ref{ratiofit1})).
For comparison, results obtained from fits to the data given 
in \cite{gi99} and \cite{ca02} are also shown.}
\label{fig5t}
\end{figure} 
The perpendicular diffusion coefficient is most easily parametrized from the results for the ratio 
$D_\perp/D_\parallel$, since this ratio exhibits little dependence with the rigidity.
In order to avoid the subdiffusive regime (see discussion below), we 
first restricted ourselves to $\rho\geq 0.03$ and $\sigma^2\geq 1$. 
As in previous numerical investigations \cite{gi99,ca02}, we found the ratio $D_\perp/D_\parallel$ to be independent 
of rigidity in the low-rigidity region (namely, for $\rho\leq 0.2$), while it scales as $\rho^{-2}$ in the high-rigidity 
region. Hence, our results can be conveniently parametrized by means of the expression
\begin{equation} 
{{D_\perp}\over{D_\parallel}}=N_\perp\times(\sigma^2)^{a_\perp}\times
\left\{\begin{array}{cl}1\ \ \ \ \ \ \ \ \ \ \ (\rho\leq 0.2)\\
\left(\rho/0.2\right)^{-2}\ (\rho>0.2)\end{array}\right.\ ,
\label{ratiofit1}
\end{equation}     
where the parameters $N_\perp$ and $a_\perp$ are also given in Table 1. In \cite{ca02a,ca02b,ca03}, a fit to the data of \cite{ca02}
(for the low-rigidity regime) was given as
\begin{equation}  
{{D_\perp}\over{D_\parallel}}={{1}\over{1+4.5^2/\left(\sigma^2\right)^{1.5}}}\ ,
\label{ratiofit2}
\end{equation}  
while the results presented in \cite{gi99}, which correspond to the turbulence range $0.03\leq\sigma^2\leq 1$, can be accounted for 
by the expression
\begin{equation}  
{{D_\perp}\over{D_\parallel}}=0.025\times(\sigma^2)^{1.835}\ .
\label{ratiofit3}
\end{equation}  

Figure 5 shows the fit given by Eq.(\ref{ratiofit1}) for the Kolmogorov case and different turbulence levels. For the sake of 
comparison, the results of \cite{ca02} (as given by Eq.(\ref{ratiofit2})) and of \cite{gi99} (as 
given by Eq.(\ref{ratiofit3})) are also shown. For moderate turbulence levels ($\sigma^2=1$) our results show an excellent agreement 
with the results of \cite{gi99}, while the agreement with \cite{ca02} tends to be better for larger turbulence.  

\begin{figure}[t!]
\centerline{{\epsfxsize=5.truein\epsfysize=3.5truein\epsffile{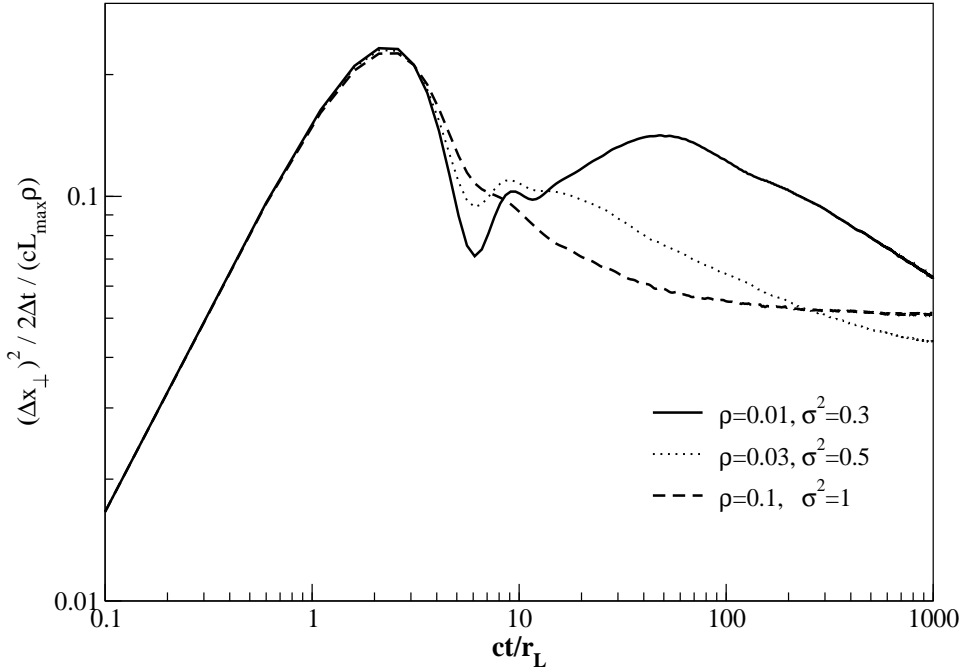}}}
\caption{$\langle\left(\Delta x_\perp\right)^2\rangle/2\Delta t/(cL_{max}\rho)$ versus time
for different values of rigidity and turbulence levels, in the case of a Kolmogorov spectrum of fluctuations.
The subdiffusive regime shows up at low enough rigidities and turbulence levels.}   
\label{fig6}
\end{figure} 
For low rigidities and turbulence levels (for instance, $\rho\simeq 0.01$ and $\sigma^2\leq 1$), we have 
found evidence for the phenomenon of subdiffusion, as already reported in \cite{ca02}.  
Indeed, a low-rigidity particle at low turbulence tends to remain attached to a field line, and hence 
transverse diffusion chiefly proceeds by the transverse random walk of field lines. The mean perpendicular displacement 
is then expected to evolve more slowly with time than in the case of normal diffusion 
(i.e., $\langle\Delta x_\perp^2\rangle\propto t^m$ with $m<1$). Figure 6 shows
$\langle\left(\Delta x_\perp\right)^2\rangle/2\Delta t/(cL_{max}\rho)$ as a function of time, 
for different values of rigidity and turbulence levels, in the case of a Kolmogorov spectrum of fluctuations. 
While a plateau is attained at large $t$ for $\rho=0.1$ and $\sigma^2=1$, 
hence corresponding to the usual diffusion relation $\langle\Delta x_\perp^2\rangle\propto t$, for
lower values of rigidity and turbulence level the phenomenon of subdiffusion shows up. 
As an approach to understanding the subdiffusive regime, it has been proposed and 
investigated the so-called compound diffusion, in which particles are assumed to be strictly tied to the field lines,
while they scatter back and forth along the lines \cite{ge63,fo77,ur77,ko00}. For the limiting case of compound diffusion,
it turns out that $m=1/2$, while in the case of three-dimensional particle transport $m$ is expected to have
a smooth dependence with rigidity and turbulence, such that $1/2\leq m(\rho,\sigma^2)<1$. Further investigations
aiming at a detailed, quantitative description of subdiffusion are currently under progress.  
 
Concerning the antisymmetric diffusion coefficient, the numerical results can be fitted by the expression
\begin{equation}
{{D_A}\over{cL_{max}\rho}}={{1}\over{3}}{{1}\over{\sqrt{1+\left(\sigma^2/\sigma^2_0\right)^2}}}\ ,
\label{danum1}
\end{equation}
where 
\begin{equation}
\sigma_0^2(\rho)=N_A\times\left\{\begin{array}{ll}\rho^{0.3}\ \ \ \ \ \ (\rho\leq 0.2)\\
1.9\ \rho^{0.7}\ (\rho>0.2)\end{array}\right.\ , 
\label{danum2}
\end{equation}
and where the values for the parameter $N_A$ are given in Table 1. In the limit of very low turbulence, this 
expression tends to the appropriate value $D_A\simeq cr_L/3$ \cite{pt93} (see also Eq.(\ref{da}) in the large
$\tau_A$ limit), while
it vanishes in the limit of very strong turbulence, as expected. 
\begin{figure}[t!]
\centerline{{\epsfxsize=5.truein\epsfysize=3.5truein\epsffile{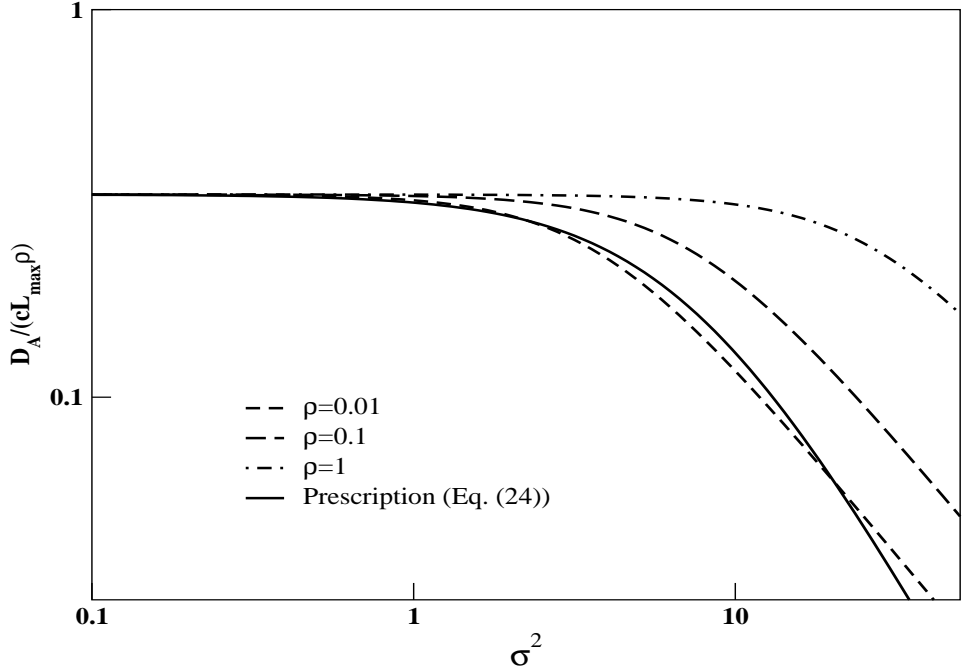}}}
\caption{Comparison between the fit to $D_A$ given in this work 
(for the Kolmogorov spectrum and the rigidities $\rho=0.01,0.1,1$) and the simple $\rho-$independent prescription 
adopted in \cite{ca02a,ca02b,ca03} (see Eq.(\ref{presc})).}
\label{fig5}
\end{figure} 
In \cite{ca02a,ca02b,ca03}, the propagation of cosmic 
rays diffusing in the Galaxy was studied and, due to the lack of an expression like that given by 
Eqs.(\ref{danum1})-(\ref{danum2})
(i.e. valid even under highly turbulent conditions), a simple $\rho-$independent prescription for $D_A/cL_{max}\rho$ 
was adopted, namely 
\begin{equation}
{{D_A}\over{cL_{max}\rho}}={{1}\over{3}}\ {{1}\over{1+\left(\sigma^2\right)^{1.5}/4.5^2}}\ ,
\label{presc}
\end{equation}
which was inferred for a Kolmogorov spectrum of fluctuations. In Figure 7 this ad hoc prescription is compared 
to the actual fit of Eqs.(\ref{danum1})-(\ref{danum2}), 
calculated for the Kolmogorov case and the rigidities $\rho=0.01,0.1,1$. 
It can be observed a sound agreement between the prescription of \cite{ca02a,ca02b,ca03} and the
fit corresponding to $\rho=0.01$, and, in any case, the dependence of $D_A/\rho$ with $\rho$ is apparent only for very
high turbulence levels $\sigma^2\gg 1$.    

As a summary, in this work we performed extensive Monte Carlo simulations to determine the parallel, transverse
and antisymmetric diffusion coefficients that describe the propagation of cosmic rays under highly turbulent
conditions. We examined the simple analytical approach proposed in \cite{bi97}, and found that the expressions given
there in terms of mean trajectory decorrelations are only meaningful for low turbulence levels. Furthermore, we 
found that, as long as their approach is valid, the decorrelation timescales associated to transverse and antisymmetric
diffusion are indeed equal. 
We evaluated the diffusion coefficients and parametrized the results by means of simple expressions, which agree 
with the expected behavior in the limit of low turbulence levels. Moreover, the results obtained were compared 
to the previous numerical calculations performed in \cite{gi99,ca02}. 
In this respect, this work extends 
the previous investigations, since it also takes into consideration other possible turbulence spectra in addition to the Kolmogorov case
(namely, the Kraichnan and the Bykov-Toptygin spectra), it provides useful parametrization formulae, and it includes
the study of the antisymmetric diffusion coefficient. Finally, we compared the new results for the antisymmetric coefficient 
with the prescription adopted previously in \cite{ca02a,ca02b,ca03} for explaining the cosmic ray spectrum, composition and
anisotropies in the region between the knee and the ankle. 
We hope that these results will be useful in a variety of different astrophysical scenarios related to the origin and transport 
of cosmic rays.

\section*{Acknowledgments}
Work partially supported by CONICET and Fundaci\'on Antorchas, Argentina. E.R. is 
partially supported by a John Simon Guggenheim Foundation fellowship. 
J.C. is currently supported by the Program for Latin American Students of the Theoretical Physics Department of Fermilab. 
Fermilab is operated by Universities Research Association Inc. under
contract no. DE-AC02-76CH02000 with the DOE.

\end{document}